\documentstyle[amsfonts,12pt]{article}
\newcommand{\be}{\begin{equation}}
\newcommand{\ee}{\end{equation}}
\newcommand{\bea}{\begin{eqnarray}}
\newcommand{\eea}{\end{eqnarray}}
\newcommand{\beas}{\begin{eqnarray*}}
\newcommand{\eeas}{\end{eqnarray*}}
\newcommand{\bi}{\begin{itemize}}
\newcommand{\ei}{\end{itemize}}
\newcommand{\bc}{\begin{center}}
\newcommand{\ec}{\end{center}}
\newcommand{\bfl}{\begin{flushleft}}
\newcommand{\efl}{\end{flushleft}}
\newcommand{\bfr}{\begin{flushright}}
\newcommand{\efr}{\end{flushright}}
\newcommand{\f}{\frac}

\def\6{\partial}  
  
\def\e{\epsilon}
  
  \def\l{\lambda}
\def\m{\mu} \def\n{\nu}  
\def\r{\rho} \def\s{\sigma} \def\t{\tau}

  \def\S{\Sigma}
  \def\O{\Omega}

\newcommand{\EE}{{\cal E}}

\begin{document}

\title{On The Symplectic Two-Form of Gravity in Terms of Dirac Eigenvalues}

\author{M. C. B. Abdalla$^{a}$\footnote{mabdalla@ift.unesp.br},
M. A. De Andrade$^{b}$\footnote{marco@cbpf.br, marco@gft.ucp.br}, 
M. A. Santos$^{c}$\footnote{masantos@gft.ucp.br} and 
I. V. Vancea$^{d}$\footnote{ivancea@unesp.ift.br, vancea@cbpf.br}}

\date{\small $a$,$d$
Instituto de F\'{\i}sica Te\'{o}rica , Universidade Estadual Paulista \\ 
Rua Pamplona 145, 015405-900.  S\~ao Paulo - SP, Brasil\\
\vspace{.5cm}
$^b$Centro Brasileiro de Pesquisas F\'{\i}sicas,\\
Rua Dr. Xavier Sigaud 150, 22290-180 Rio de Janeiro - RJ, Brasil\\
\vspace{.5cm}
$^c$Departamento de F\'{\i}sica, Universidade Federal Rural do Rio de Janeiro\\
23851-180, Serop\'{e}dica - RJ, Brasil}

\maketitle

\abstract{The Dirac eigenvalues form a subset of observables of the Euclidean 
gravity. The symplectic two-form in the covariant phase space could be 
expressed, in principle, in terms of the Dirac eigenvalues. We discuss 
the existence of the formal
solution of the equations defining the components of the symplectic form in 
this framework.}

\newpage

One of the major obstacles in quantizing the gravity is finding a complete set
of observables of it. Recently, a certain progess has been made in defining 
and manipulating covariant observables in various theories of gravity
\cite{piro,pero,ro1,ro2}. Previous works
showed that the Dirac eigenvalues can be considered as observables of
gravity, too, on manifolds endowed with an Euclidean structure 
\cite{laro1,laro2,la}. The result was generalized to include the local
$N=1$ supersymmetry in \cite{va1,va2,va3,va4} and it was shown to be 
connected to spectral geometry in \cite{vac}. However, in order to completely
understand the covariant phase space of the Euclidean gravity in terms of the
Dirac eigenvalues, one has to know what is the form of the symplectic two-form
in terms of the observables. The aim of this letter is to discuss the 
existence of a formal solution of this problem. 

Let us begin by considering a four-dimensional compact manifold $M$ without 
boundary endowed with an Euclidean metric field $g_{\m\n}$. One introduces
a tetrad field which maps the metric at each point $x \in M$ to the 
local Euclidean  metric in the tangent space: 
$g_{\m\n}(x)=E_{\m}^{I}(x)E_{\n}^{J}(x)\delta_{IJ}$. The covariant phase space 
of the theory is given 
by non-equivalent solutions of the Eistein equations on $M$ modulo the
``gauge transformations'', i. e. transformations generated by local 
$SO(4)$ times diffeomorphisms. The functions of the phase space are observables
of the theory. Consider now the Dirac equation 
\begin{equation}
D \left | \right . \psi_n \rangle = \l_n \left | \right .\psi_n \rangle ,
\label{deq}
\end{equation}
where $\left| \psi_n \right\rangle$ is a spinor field (in the Dirac's bra-ket 
notation) and $n$ is a positive integer (for simplicity, we assume that the 
Dirac operator $D$ has no zero eigenvalue.) The eigenvalues $\l_n$ 
define a discrete family of real valued functions on the space of smooth 
tetrads $\EE$ and a function from $\EE$ into the space of infinite 
sequences $R^{\infty}$
\bea
{\l_{n}} &:& \EE \longrightarrow R ~~~~,~~E \rightarrow
{\l_{n}}[E], \\ \label{funct}
{\l_{n}} &:& \EE \longrightarrow R^{\infty}~~,~~E \rightarrow
\{ {\l_{n}}[E] \}. \label{seq}
\eea
For every $n$, $\l_n[E]$ is invariant under the gauge 
group action on the tetrads \cite{laro1}. In general, $\l_n$ do not form a
set of coordinates neither on the space of gauge orbits nor on the phase space
\cite{laro1}.

In order to analyse the phase space further, one has to define the Poisson 
structure on the set of the eigenvalues. This can be achieved by constructing
firstly the symplectic two-form of general relativity \cite{abr}  
\be
\O (X,Y) = \f{1}{4}\int_{\S} d^3 \sigma n_{\r}
[X^{a}_{\m} , \stackrel{\leftrightarrow}{\nabla_{\t}} Y^{b}_{\n} ]
\e^{\t}_{ab\upsilon } \e^{\upsilon \r \m \n} , 
\label{sform}
\ee
where $X^{a}_{\m}[E]$ define a vector field on the phase space and the
brackets are given by
\be
[X^{a}_{\m} , \stackrel{\leftrightarrow}{\nabla_{\t}} Y^{b}_{\n} ] =
X^{a}_{\m} \nabla_{\t} Y^{b}_{\n} - 
Y^{a}_{\m} \nabla_{\t} X^{b}_{\n} . \label{brack}
\ee
Here, $\S$ is an arbitrary Arnowitt-Deser-Misner surface an $n_{\r}$ is its
normal one form. The two-form $\O$
is invertible only on the space of gauge fixed fields since it is degenerate
on the space of the solutions of the Einstein equations. The coefficients
of $\O$ are given by the following relation
\begin{equation}
\O^{\m\n}_{IJ}(x,y) = \! \int_{\S}\! d^{3}\sigma \ n_{\rho}\  
[\delta(x,x(\sigma)) \overleftarrow{\overrightarrow{\nabla}}_{\tau}  
\delta(y, x(\sigma)) ]\ \epsilon^{\tau}{}_{IJ\upsilon} 
\,\epsilon^{\upsilon\rho\mu\nu} .
\label{Ocoeff}
\end{equation} 
As was already noted in \cite{laro2}, the symplectic two-form (\ref{sform})
can be written in terms of the Dirac eigenvalues if the map (\ref{funct}) is
locally invertible on the phase space. Then, the coefficients $\O_{mn}$ of 
$\O$ defined by the following relation
\be
\O = \O_{mn} d\l_n \wedge d\l_m,
\label{Ocoeffl}
\ee
can be expressed in terms of (\ref{Ocoeff}) as follows 
\begin{equation}
\O_{mn}\ T_{n}{}^{\mu}_{I}(x)\ T_{m}{}^{\nu}_{J}(y)= 
\O^{\m\n}_{IJ}(x,y),
\label{Equation}
\end{equation}
where 
\begin{equation}
T_{n}{}^{\mu}_{I}(x) = \frac{\delta \l_{n}[E]}{\delta E_{\mu}^{I}(x)}. 
\label{Tdef}
\end{equation}

In order to have a complete description of the phase space of the theory in 
terms of the Dirac eigenvalues, one has to express the coefficients 
$\O_{mn}$ in terms of $\O^{\m\n}_{IJ}(x,y)$, that is to invert (\ref{Tdef}).
To this end, we introduce the following objects which are well defined since
the map (\ref{funct}) is invertible (a necessary condition for the
existence of $\O_{mn}$)
\be
U_{n}{}^{I}_{\m}(x)= \frac{\delta E^{I}_{\m}(x)}{\delta \l_n}.
\label{Udef}
\ee
A simple algebra shows that the following two relations hold
\bea
U_{n}{}^{I}_{\m}(x)T_{n}{}^{\n}_{j}(y)&=&\delta^{I}_{J}
\delta^{\n}_{\m}\delta^{(4)}(x-y),\label{invIJ}\\
U_{n}{}^{I}_{\m}(x)T_{m}{}^{\mu}_{I}(x)&=&\delta_{mn}.\label{invmn}
\eea
Note that the coefficients $\O_{mn}$ do not depend explicitely on the point 
$x$ of $M$. Moreover, since the eigenvalues $\l_n$ of $D$ are defined globally
on $M$, $\O_{mn}$ has the same property. Therefore, in order to elliminate the
dependence on the points $x$ and $y$, one has to integrate twice over $M$ when
inverting (\ref{Equation}).
Then, using (\ref{Udef}) and (\ref{invmn}) one can obtain from (\ref{Equation})
the following relation
\be
\O_{mn}=\frac{1}{V_M^2}\int_{M}Dx\int_{M}Dy ~ \O^{\m\n}_{IJ}(x,y)
~ U_{n}{}^{I}_{\m}(x) ~ U_{m}{}^{I}_{\n}(y),
\label{Inverse}
\ee
where $Dx = d^4x\sqrt{g}$ 
and $V_M$ is the four-volume of $M$.
Note that in order to obtain the relation (\ref{Equation}) from 
(\ref{Inverse}) one has either to rescale the relation (\ref{invmn}) by a 
factor of $V_M$ in the r.h.s. or to rescale the delta-function integral on $M$.
In what follows, we are going to use the relation
\be
\int_{M}Dx~f(x)\delta^{(4)}(x-y) = V_M f(y).
\label{Delta}
\ee 

A formal solution of (\ref{Inverse}) can be given once  
$U_{n}{}^{I}_{\m}(x)$'s are known. To calculate them, we use the Dirac
equation (\ref{deq}). Assume that the Dirac eigenspinors satisfy the
global orthoganality and closure relations
\bea
\langle \psi_n \left | \right. \psi_m \rangle &=& \delta_{nm},
\label{globalorth}\\
\sum_n \left | \right. \psi_n \rangle \langle \psi_n \left | \right. &=& 
{\Bbb I}, \label{closure}
\eea
where the scalar product in the Hilbert space of the vector fields on $M$ is
defined as 
\be
\langle \psi \left | \right. \phi \rangle = \int_{M} Dx~
\langle \psi (x) \left | \right. \phi (x) \rangle .   
\label{globalscal}
\ee
Here, $\langle \psi (x) \left | \right. \phi (x) \rangle $ is the scalar 
product in the local spinor fiber $S_x(M)$ over $x$. The local spinor 
sections
$ \{ \left | \right. \psi_n (x) \rangle \} $ are induced by the 
fields
$ \{ \left | \right. \psi_n \rangle \}$. We assume further that the global 
fields are defined by integral of local spinors
\be
\left | \right. \psi_m \rangle  =  \int_{M} Dx ~ 
\left | \right. \psi_m (x) \rangle .
\label{intloc} 
\ee
It is worth to notice that the relation above implies the following bilocal 
action of the Dirac operator
\be
D(x)\left | \right. \psi_m (y) \rangle = V_M^{-1}\delta^{(4)}(x-y)
\left | \right. \psi_m(x) \rangle ,
\label{localeq}
\ee
where the relation (\ref{Delta}) was taken into account.
Then, the bilocal scalar product and the local closure relations are given by
the following relations
\bea
\langle \psi_n (x) \left | \right. \psi_m (y)\rangle &=& V_{M}^{-1}\delta_{nm}
\delta^{(4)}(x-y),
\label{localorth}\\
\sum_n \left | \right. \psi_n (x) \rangle \langle \psi_n (x)\left | \right. 
&=& V_{M}^{-1}{\Bbb I}. 
\label{locclosure}
\eea
The local orthogonality and closure relations must be defined in order to deal
with the local terms in the relation (\ref{Inverse}). 

The next step is to project $\O_{mn}$ onto the basis formed by the 
eigenspinors of $D$. Since the coefficients of the symplectic form in the
basis formed by $\l_n$ are globally defined on $M$, the projection should 
be performed onto the basis $ \{  \left | \right. \psi_n \rangle \} $ 
rather than onto $ \{  \left | \right. \psi_n (x) \rangle \} $. By
using the relations (\ref{globalorth}), (\ref{closure}), (\ref{intloc}),
(\ref{localorth}) and (\ref{locclosure}), one can easily show that the 
components of $\O_{mn}$ are given by
\bea
[\O_{mn}]_{st} = V_{M}^{2} \int Dx \sum_{r,k} [\O^{\m\n}_{IJ}(x,x)]_{sr}
[U_{n}{}^{I}_{\m}(x)]_{rk}[U_{m}{}^{j}_{\n}(x)]_{kt},
\label{compO}
\eea 
where we are using the following shorthand notations 
\be
[\O_{mn}]_{st} = \langle \psi_s  \left | \right. [\O_{mn}] 
\left| \right. \psi_t \rangle , \hspace{0.5cm}
[\O^{\m\n}_{IJ}(x,y)]_{sr} =
\langle \psi_s (x) \left | \right.  [\O^{\m\n}_{IJ}(x,y)] 
\left| \right. \psi_r (y)\rangle ,
\label{notations1}
\ee
\be
[U_{n}{}^{I}_{\m}(x)]_{rk} = 
\langle \psi_r (x) \left | \right.  
[U_{n}{}^{I}_{\m}(x)] 
\left| \right. \psi_k (x)\rangle .
\label{notations2}
\ee
Given the manifold $M$, one could calculate, in principle, the matrix
elements of $[\O^{\m\n}_{IJ}(x,x)]$ after computing the spectrum of the
Dirac operator. What is left are the matrix entries from (\ref{notations2}).
To obtain them we derive the local Dirac equation with respect to the 
eigenvalue $\l_m$. The resulting relation has the following form
\be
V_{M}^{2}\sum_{k}[U_{m}{}^{I}_{\m}(x)]_{rk}[D^{\m}_{I}(x)]_{kn} = 
\delta_{mn}\delta_{rn},
\label{Diracelem}
\ee
for all $m$,$r$ and $n$. Here, the sum is over $k$ only and
\be
[D^{\m}_{I}(x)]_{kn} =  
\langle \psi_k (x) \left | \right.  
[D^{\m}_{I}(x)] 
\left| \right. \psi_n (x)\rangle .
\label{notations3}
\ee
These terms are determined by the eigenspinors of $D$ and by noting that
\bea
D^{\m}_{I}(x) &=& i \gamma^J \{ -E^{\n}_{I}(x)E^{\m}_{J}(x)(\6_{\n} +
\omega_{\n KL}(x)\sigma^{KL}) \nonumber\\
&-&\frac{1}{2}E^{\n}_{J}(x)[
E^{\rho}_{I}(x)E^{\mu}_{K}(x)(\6_{\n}E_{\rho L}(x) -
\6_{\rho}E_{\n L}(x)) + \delta_{LI}\6_{\s}(E^{\m}_{K}(x)\delta^{\s}_{\n}-
E^{\s}_{K}(x)\delta^{\m}_{\n})\nonumber\\
&+&
\delta^{M}_{I}\delta^{\m}_{\n}E^{\tau}_{K}(x)E^{\sigma}_{L}(x) 
- E^{\tau}_{I}(x)E^{\mu}_{K}(x)E^{\sigma}_{L}(x)E^{M}_{\n}(x)
\nonumber\\
&-& E^{\tau}_{K}(x)E^{\s}_{I}(x)E^{\mu}_{L}(x)E^{M}_{\n}(x))
\6_{\s}E_{\t M}(x) \nonumber\\
&-&\6_{\rho}(E^{\mu}_{K}(x)E^{\rho}_{L}(x)E_{\n I}(x))
- (K \leftrightarrow L)]\sigma^{KL} 
\},
\label{derivD}
\eea
where $\omega_{\m IJ}(x)$ are the components of the spin-connection in the 
spin bundle $S(M)$ over $M$, $\gamma^I$ are the tangent-space Dirac matrices 
and $\sigma^{IJ}=\frac{1}{4}[\gamma^I,\gamma^J]$. In principle, one can 
compute the matrix elements of (\ref{derivD}) if the vielbein is fixed and 
the eigenspinors of $D$ are known. Therefore, one has to solve the 
system (\ref{Diracelem}) in order to find $[\O_{mn}]_{st}$. 

In general, the Dirac operator may have an infinite set of eigenspinors on 
$M$, which makes the system (\ref{Diracelem}) infinite, too. A necessary and 
sufficient condition for the  determinant 
${\mbox{det}} [D^{\m}_{I}(x)]_{kn}$ be 
absolutely convergent \cite{sb} is that the product 
$\prod_{k}|[D^{\m}_{I}(x)]_{kk}|$
converges absolutely and there is a non-negative integer number $p$ such that
\be
\sum{k}[\sum_{l}|[D^{\m}_{I}(x)]_{kl}|^{p}]^{\frac{1}{p-1}}~,~~k\neq l
\label{conv}
\ee
be convergent, too. Let us assume that this is the case. Then the solution
of the system (\ref{Diracelem}) is determined by the convergence of the
determinant of $[D^{\m}_{I}(x)]_{kn}$. If it converges to zero, the elements 
of the matrices  $ V_{M}^{2}[U_{m}{}^{I}_{\m}(x)]$ may exist, but are 
undetermined. A more interesting case is given by a non-zero determinant of
$[D^{\m}_{I}(x)]_{kn}$. Then the inverse of this matrix can be constructed.
If we put $m=n$ in (\ref{Diracelem}), we see that some of the elements of the
full set of matrices  $\{ V_{M}^{2}[U_{m}{}^{I}_{\m}(x)] \}$ are proportional
to the elements of the inverse of $[D^{\m}_{I}(x)]_{kn}$. Actually, the
condition $m=n$ determines the following elements of the matrices
$\{ V_{M}^{2}[U_{m}{}^{I}_{\m}(x)] \}$ 
\be
[U_{n}{}^{I}_{\m}(x)]_{nk} = V_M^{-2}[D^{\m}_{I}(x)]^{-1}_{nk},
\label{solution1}
\ee
for all $k$. Here, $[D^{\m}_{I}(x)]^{-1}_{kn}$ represent the elements of 
the inverse of $[D^{\m}_{I}(x)]_{kn}$. These elements exist if the 
complement of each element of $[D^{\m}_{I}(x)]$ converges. The condition that
$[U_{n}{}^{I}_{\m}(x)]_{nk}$ form a matrix (more exactly, the condition that
the matrix product between $[U_{n}{}^{I}_{\m}(x)]_{nk}$ and the inverse
$[D^{\m}_{I}(x)]^{-1}_{nk}$ be well defined) leads to the following 
supplementary relations:
\be
[U_{s}{}^{I}_{\m}(x)]_{sk}=
[U_{1}{}^{I}_{\m}(x)]_{sk}=
[U_{2}{}^{I}_{\m}(x)]_{sk}
=\cdots ,
\label{solution2}
\ee
for all $k$. The relations (\ref{solution1}) and (\ref{solution2}) show that
the set of the matrices  $\{ V_{M}^{2}[U_{m}{}^{I}_{\m}(x)] \}$ is degenerate
and that these matrices should be proportional to the inverse of the 
matrix of expected values of the Dirac operator. In the case when $m \neq n$ 
the
r.h.s of (\ref{Diracelem}) vanishes. Therefore, if the determinant of 
$[D^{\m}_{I}(x)]_{kn}$ does not vanishes, the corresponding matrix elements
should be zero. We note that the two cases are incompatible with each other,
since the same matrix elements of  $ V_{M}^{2}[U_{m}{}^{I}_{\m}(x)]$ enter
both of them and, while in the case $m=n$ they are determined in terms of
the elements $[D^{\m}_{I}(x)]^{-1}_{nk}$, in the case $m \neq n$ they are 
either zero or are undetermined.

To conclude, one can formally solve the equation (\ref{Equation}) as in
(\ref{Inverse}). If the determinant of $[D^{\m}_{I}(x)]_{kn}$ is different 
zero, the terms $[U_{m}{}^{I}_{\m}(x)]_{kl}$ from (\ref{Inverse}) are
determined by the relations (\ref{solution1}) and (\ref{solution2}). 
The matrices $\{ V_{M}^{2}[U_{m}{}^{I}_{\m}(x)] \}$ should be all proportional
to the inverse $[D^{\m}_{I}(x)]^{-1}$ and the solution is degenerate in 
this sense.  It is not clear 
if this degeneracy is related in some way to the fact that we are dealing 
with the set of smooth vielbeins instead of the gauge fixed ones.

{\bf Acknowledgments}

M. A. S. and I. V. V. would like to thank to J. A. Helayel-Neto for 
hospitality at CBPF during the preparation of this work. I. V. V. would like 
to acknowledge to N. Berkovits and O. Piguet for useful discussion and to 
FAPESP for support within a postdoc fellowship.

\end{document}